\documentclass[9pt,twocolumn,twoside]{pnas-new}
\usepackage{lipsum}
\usepackage{bm}                                 % bold math
\usepackage{mathtools}                          % tools to create operators and delimiters
\usepackage{subcaption}
\usepackage{siunitx}

\usepackage{xr}
\makeatletter

\newcommand*{\addFileDependency}[1]{% argument=file name and extension
\typeout{(#1)}% latexmk will find this if $recorder=0
\@addtofilelist{#1}
\IfFileExists{#1}{}{\typeout{No file #1.}}
}\makeatother

\newcommand*{\myexternaldocument}[1]{%
\externaldocument[si-]{#1}%
\addFileDependency{#1.tex}%
\addFileDependency{#1.aux}%
}
\myexternaldocument{SI/SI}

\templatetype{pnasresearcharticle} % Choose template 
\title{Automated Design of Tubular Origami with Anisotropic Stiffness}

\author[a]{Mingkai Zhang}

\author[a,b]{Davood Farhadi}

\affil[a]{Precision and Microsystems Engineering, Delft University of Technology, The Netherlands}

\affil[b]{John A. Paulson School of Engineering and Applied Sciences, Harvard University, Cambridge, MA 02138, USA}

\leadauthor{Zhang}

\authorcontributions{}
\authordeclaration{The authors declare no competing interest.}
\correspondingauthor{\textsuperscript{1}dfarhadim@gmail.com}


\begin{abstract}
Thin sheets can be assembled into tubular origami structures that combine deployability with pronounced anisotropic stiffness, enabling applications ranging from robotics to deployable systems. However, most existing tubular origami designs remain limited to degree-four vertex topologies and are characterized primarily in axial and radial loading modes, without a full assessment of anisotropic stiffness. Here, we present an automated design framework for tubular origami that jointly explores local vertex topology through generalized degree-$n$ vertices and global tube topology through the polygonal cross-section, for the systematic design and optimization of anisotropic stiffness. Using a calibrated bar-and-hinge model together with experimental validation, we quantify large-deformation stiffness responses in axial translation, in-plane translation, torsion about the tube axis, and rotation about in-plane axes, thereby characterizing the anisotropic stiffness of the tube across its compliant and constrained deformation modes. The resulting design-space exploration showed that the polygonal cross-sectional topology is the primary factor governing the anisotropic stiffness. We further show that increasing the local vertex degree can improve global structural performance, particularly for tubes with a small number of cross-sectional vertices, demonstrating that higher local kinematic freedom does not necessarily compromise stiffness at the structural scale. Compared with a benchmark design, the optimized architectures achieve more than 50 times higher constrained rotational stiffness. Together, these results highlight higher-degree vertices and polygonal cross-sectional topology as powerful design variables for tailoring anisotropic stiffness in tubular origami.
\end{abstract}

\dates{This manuscript was compiled on \today}

\begin{document}

\maketitle
\thispagestyle{firststyle}
\ifthenelse{\boolean{shortarticle}}{\ifthenelse{\boolean{singlecolumn}}{\abscontentformatted}{\abscontent}}{}

\section{Introduction}
\label{sec:introduction}
Origami, a paper-folding technique, has emerged as a powerful design paradigm in science and engineering due to its ability to transform two-dimensional (2D) sheets into complex three-dimensional (3D) structures with reconfigurable functionalities \citep{overvelde2016-B-circle,overvelde2017-B-block,chen2015origami,novelino2020untethered,chai2025design,sharma2025multi,xia2025deployment, nelson2019origami}. Through the strategic programming of crease patterns, origami structures can exhibit diverse mechanical behaviors, including stiffness modulation \citep{zhai2018origami-tunnablestiff,liu2018topological,junfeng2024modular}, programmable elastic modulus \citep{silverberg2014using-progamE,sengupta2018harnessing}, multistability \citep{melancon2022inflatable, yasuda2017origami-JK-nc-bistable,silverberg2015origami,waitukaitis2015origami}, negative Poisson’s ratio responses \citep{yasuda2015-JK-negP,wang2020modulation}, energy absorption \citep{zhang2023energy,he2024energy,qiang2024energy}, and wave control \citep{fang2026manipulating,li2025double}. These properties have enabled origami-inspired systems across space exploration \citep{morgan2016approach-ori-aerospace,sigel2014application,zirbel2013accommodating}, architecture \citep{melancon2021multistable, del2010adaptive-ori-build,buri2008origami,reis2015transforming}, robotics \citep{farhadi2025origami, ze2022soft, rus2018design-ori-robots,onal2011towards,onal2012origami}, and biomedical devices \citep{randall2012self-oribio,amir2014universal,johnson2017fabricating,coles2024origami}.

Among the many origami topologies, tubular origami structures are distinguished by their pronounced anisotropic mechanical behavior, combining deployability along the axial direction with high stiffness in radial and transverse directions. This combination has enabled applications in robotic locomotion \citep{onal2011towards,onal2012origami}, continuum robotic arms \citep{zhang2022design}, biomedical stents \citep{kuribayashi2006self-tubeBio}, and surgical tool support structures \citep{sargent2020origami}. Beyond applications, tubular origami offers exceptional stiffness-tuning capabilities, including large contrasts between in-plane and out-of-plane stiffness \citep{cheung2014origami-miru-block-stiff} and programmable translational stiffness through geometric coupling and stacking strategies \citep{filipov2015origami-pnas-tube,lin2020folding-couple-tube-block}.

From a topological perspective, most tubular origami designs are derived from a limited set of vertex configurations, including Miura-ori \citep{miura1994map}, Kresling \citep{masana2019equilibria}, and Waterbomb patterns \citep{hanna2014waterbomb,chen2016symmetric}, corresponding to a particular degree-four, degree-six, and degree-eight vertices, respectively. Degree-four vertices have received the most attention due to their limited kinematic freedom, which simplifies design while enabling reconfigurability and anisotropic stiffness. Such topologies allow polygonal and translationally symmetric cross-sections to morph between multiple geometries \citep{filipov2016origami-cross-section}. Extensions include star-shaped cross-sections \citep{kamrava2019origami-star} and heterogeneous tubular structures capable of transitioning between distinct unit-cell configurations \citep{miyazawa2021heterogeneous-reshape-trajectory}. However, these developments have largely remained confined to low-degree vertex configurations.

Despite these advances, two key challenges remain. First, existing studies predominantly focus on axial and radial stiffness, while rotational stiffness—essential for a complete characterization of anisotropic mechanical behavior—has received limited attention. Second, exploration of tubular origami has remained biased toward degree-four vertices, not due to inherent mechanical limitations, but due to the absence of systematic and scalable approaches capable of generating and evaluating the expanded design space introduced by generalized degree-$n$ vertices.

To address these challenges, this paper introduces a generalized and automatable design framework for tubular origami based on degree-$n$ vertices. The framework enables systematic generation, evaluation, and optimization of tubular architectures across vertex degree and geometric parameters, and quantifies stiffness in axial, radial, and rotational directions under large deformations. Using this approach, we demonstrate that increasing local kinematic freedom at the vertex level can counterintuitively lead to stiffness amplification and enhanced anisotropy at the tubular level through geometric coupling.

%In this paper, we first propose a design generation method for tubular origami structures based on multi-degree origami vertices, which follow a generalized cross-section characterized by a number of intersections. Objective functions are formulated to evaluate translational and torsional stiffness across all directions in three-dimensional space under large deformations. These functions subsequently derive the Pareto front, enabling assessment of trade-offs between translational and torsional performance. Next, we examine how varying both the number of degrees of the origami vertices and the number of intersections influences the optimization results. Finally, we validate our findings through experimental verification.

\section{Methods}
We first outline a generic methodology for designing tubular origami. The method automatically generates tubular origami structures composed of multi-degree vertices and applies geometric constraints to ensure feasible and deployable configurations. Translational and torsional stiffness in all spatial directions are then computed through numerical analysis. Two objective functions are formulated to capture the stiffness characteristics, forming the basis of a multi-objective optimization framework used to identify optimal designs. The resulting structures are subsequently validated through experimental testing.

\begin{figure*}[b!]
    \centering
    \includegraphics[width=\linewidth]{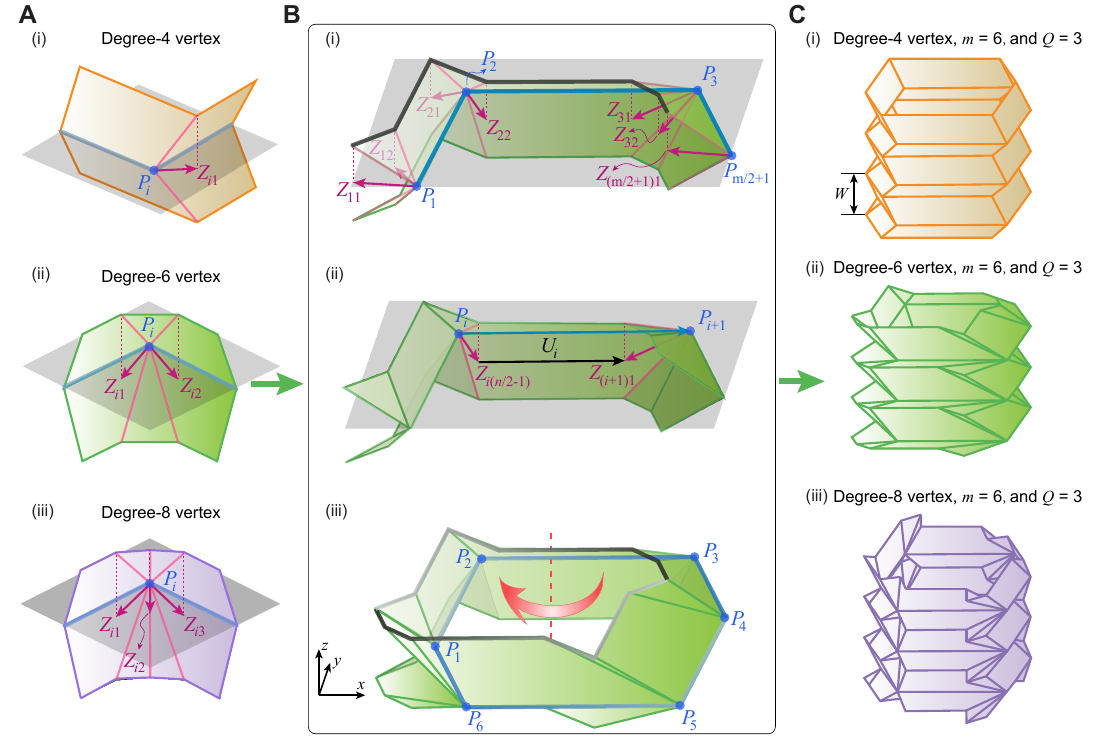}
    \caption{\textbf{Automated generation of tubular origami with general degree-$n$ vertices.} (A) Mirror-symmetric parameterization of individual origami vertices with different degrees, (i) degree-4 vertex, (ii) degree-6 vertex, and (iii) degree-8 vertex. (B) Construction of a single closed-loop origami layer from a selected degree-$n$ vertex, shown here for a degree-6 vertex as a representative example. (i) Arrangement of vertices $P_i$ along a planar polygonal chain defining the open fundamental domain prior to closure. (ii) Geometric constraints imposed to ensure the existence and compatibility of all faces. (iii) Closure of the origami loop through rotational symmetry operations, resulting in a single tubular layer. (C) Generation of tubular origami structures by stacking the closed-loop layers. Examples are shown for degree-4, degree-6, and degree-8 vertices (with $m=6$ and $Q=3$), illustrating how the same automated procedure produces tubular architectures with distinct geometries.}
    \label{fig:generation}
\end{figure*}

\subsection{Automated generation of tubular origami}
To enable automated generation of tubular origami topologies, we begin by introducing a mirror-symmetric parameterization of individual origami vertices. As illustrated in Fig.~\ref{fig:generation}A(i--iii), representative degree-4, degree-6, and degree-8 vertices are defined such that a horizontal plane passing through the vertex center acts as a symmetry plane. In each case, two crease lines lie within this plane (highlighted in blue), while the remaining crease lines occur in mirror-symmetric pairs above and below the plane (highlighted in red). This construction naturally generalizes to vertices of degree $n$. Because all non-central creases occur in symmetric pairs, the vertex degree $n$ must be even. For a general degree-$n$ vertex located at position $P_i$, the crease geometry is specified by a set of vectors $Z_{i,j}$, with $j = 1,\ldots,n/2-1$, corresponding to the orthogonal projections of the non-central crease lines onto the symmetry plane. These projected vectors define the in-plane orientations and lengths of the associated three-dimensional crease lines, while mirror symmetry uniquely determines their out-of-plane counterparts.

We next arrange multiple degree-$n$ vertices along a planar polygonal chain, whose vertices $P_i, i = 1,\ldots,m$ define the spatial locations of the origami vertices in a tubular configuration, where $m$ denotes the number of vertices in the final closed configuration. At each vertex $P_i$, the two central creases of the corresponding degree-$n$ vertex are aligned with the adjacent edges of the polygonal chain. Fig.~\ref{fig:generation}B(i) illustrates this construction for degree-6 vertices by showing the open fundamental domain, which consists of the subset of vertices $P_i$ with $i = 1,\ldots,m/2+1$ prior to closure. Once the vertex positions $P_i$ and the projected vectors $Z_{i,j}$ are specified, the upper polygonal chain and its mirror-symmetric counterpart are uniquely determined. Adjacent vertices $P_i$ and $P_{i+1}$ define a quadrilateral facet connecting the corresponding upper and lower chains. To ensure that these facets are geometrically admissible, additional constraints are imposed on the choice of the vectors $Z_{i,j}$. In particular, the vectors $Z_{i,n/2-1}$ and $Z_{i+1,1}$, which correspond to the last crease of vertex $P_i$ and the first crease of vertex $P_{i+1}$, respectively, must be arranged such that the resulting quadrilateral facet is consistently oriented and does not self-intersect (Fig.~\ref{fig:generation}B(ii)). In practice, this condition is enforced by requiring that the vector$\mathbf{U}_i = a\,\overrightarrow{P_iP_{i+1}}, \qquad a>0,$ defines a consistent orientation between adjacent facets. This constraint serves as a geometric admissibility condition ensuring the physical realizability of the quadrilateral faces. To form a closed tubular origami, we apply a rotational symmetry operation of angle $\pi$ counterclockwise about the central axis to the polygonal chain and the associated upper and lower chains (Fig.~\ref{fig:generation}B(iii)). Through this operation, the open fundamental domain is replicated and closed into a loop. The central creases of all vertices collectively trace a closed polygon, shown in blue in Fig.~\ref{fig:generation}B(iii). The number of vertices of this polygon, denoted by $m$, together with their spatial coordinates, determines the global topology of the origami loop. Fig.~\ref{fig:generation}B(iii) shows a representative example with $m=6$.

Fig.~\ref{fig:generation}C(i--iii) demonstrates that, for a fixed closed-loop polygon, selecting degree-4, degree-6, or degree-8 vertices leads to distinct tubular origami topologies. These loops can be stacked along the axial direction to form extended tubular structures. To fully specify the three-dimensional geometry of the resulting origami, two additional parameters are introduced: the number of stacked layers $Q$ and the height of each layer $W$.

Taken together, this construction defines a systematic design framework for generating tubular origami based on degree-$n$ vertices. By varying the vertex degree $n$, the number of vertices $m$ and their spatial coordinates $P_i$ defining the closed-loop polygon, the projected crease vectors $Z_{i,j}$ specifying local crease geometry, as well as the stacking parameters $Q$ and $W$, a broad family of tubular origami topologies and three-dimensional configurations can be generated within the same geometric framework.

\subsection{Fabrication and experimental characterization}
\begin{figure}[!ht]
    \centering
    \includegraphics[width=\columnwidth]{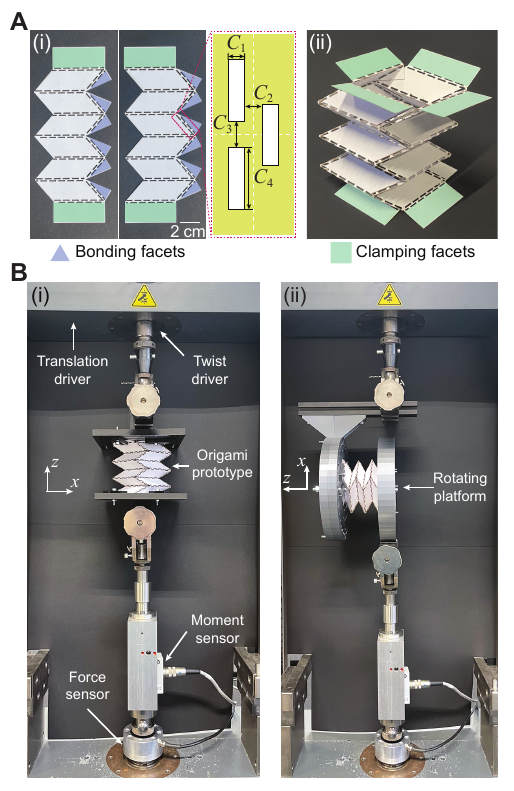}
    \caption{\textbf{Fabrication and Experimental characterization.} (A) Fabrication of tubular origami (i) laser-cut open-loop origami patterns with bonding facets (blue) and perforated crease lines. (ii) Assembly of two open-loop patterns into a closed tubular origami structure. (B) Experimental setups used for mechanical characterization. (i) setup for measuring axial translational stiffness and torsional stiffness about the $z$ axis. (ii) Modified setup incorporating a rotating platform, enabling measurements of shear stiffness along arbitrary directions within the $x$--$y$ plane and bending stiffness about axes in this plane.}
    \label{fig:setup}
\end{figure}

In this section, we describe the methodologies used to fabricate and experimentally characterize tubular origami structures. These experiments provide quantitative measurements of the mechanical response of the structures. The procedures are illustrated using a representative tubular origami generated from a degree-4 vertex.

\paragraph{Fabrication.} 

\begin{figure*}[b!]
    \centering
    \includegraphics[width=\linewidth]{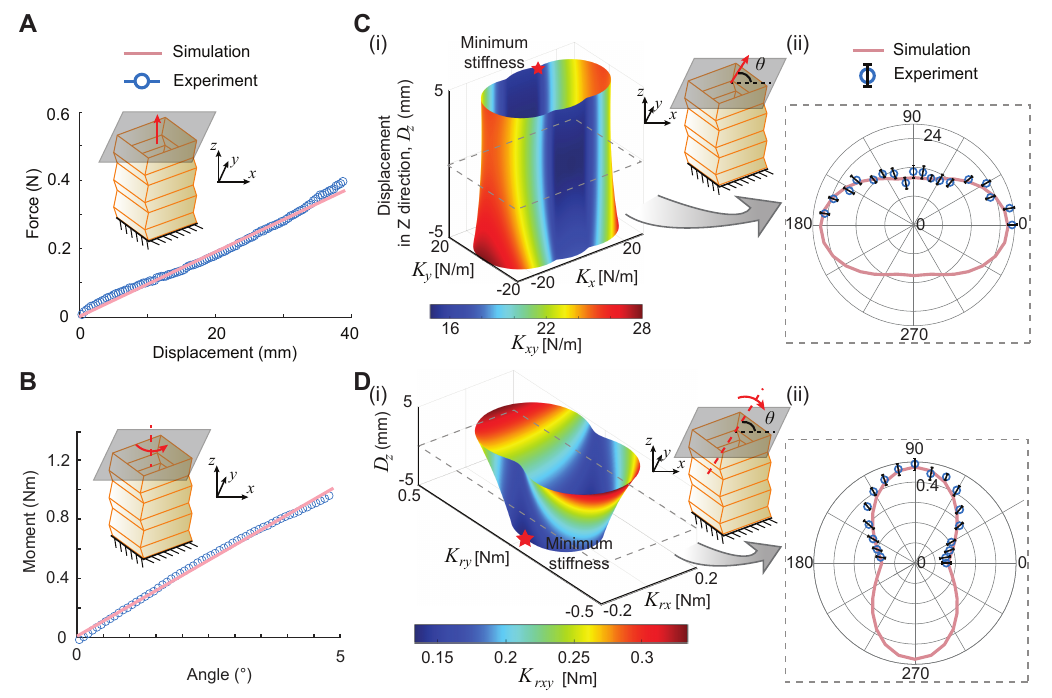}
    \caption{\textbf{Numerical stiffness analysis under large deformation.} Numerical and experimental results for (A) Translational force–deflection behavior along the z-axis, the intended compliant direction of the structure. (B) Moment–angular displacement response about the z-axis, corresponding to a stiff direction. (C) Translational stiffness in the x–y plane. (i) Directional translational stiffness measured over all in-plane orientations. The stiffness components $K_x$ and $K_y$ are evaluated at multiple prescribed displacements along the z-axis, revealing the evolution of the three-dimensional constraint stiffness as the structure deforms along its compliant direction.  (ii) Magnitude of the in-plane translational stiffness as a function of orientation angle at zero z-displacement. (D) Rotational stiffness about axes lying in the x–y plane. (i) Directional rotational stiffness components $K_{rx}$ and $K_{ry}$ obtained by rotating the axis within the x–y plane and evaluated at different z-displacements. (ii) Magnitude of the in-plane rotational stiffness at zero z-displacement as a function of the axis orientation in the x–y plane.}
    \label{fig:exp_verify}
\end{figure*}
We demonstrate the fabrication process using a representative tubular origami structure generated from a degree-4 vertex, hereafter referred to as Design I. In this design, the closed-loop polygon consists of $m=4$ vertices with positions $P_i = (x_i,y_i)$, $i = 1,\ldots,4$. The vertex coordinates are defined as
$
P_1 = (0,0), \ 
P_2 = (35.4,35.4), \ 
P_3 = (70.8,0), \ 
P_4 = (35.4,-35.4),
$
where the numerical values are chosen to prescribe the desired closed-loop geometry. For each vertex $P_i$, the local crease geometry is specified by a projected vector $Z_{i,1}$, which defines the orientation and length of the non-central crease lines. In the fabricated Design I, these vectors are identical for all vertices and are given by
$
Z_{1,1}=Z_{2,1}=Z_{3,1}=Z_{4,1}=(20,0),
$
with all dimensions specified in millimetres.

Because the resulting tubular origami geometry can be non-Euclidean, it cannot, in general, be fabricated from a single planar sheet. Instead, the closed-loop pattern is decomposed into two open-loop patterns, which are fabricated separately and subsequently assembled. Each open-loop pattern is laser-cut from flat cardboard sheets and incorporates additional bonding facets, indicated by blue triangular markers in Fig.~\ref{fig:setup}A(i). These bonding facets enable the two open patterns to be connected, resulting in a closed tubular origami structure, as shown in Fig.~\ref{fig:setup}A(ii).

The fabricated tubular origami consists of three stacked closed-loop layers ($Q=3$), each with a layer height $W = \SI{30}{mm}$. To enable controlled folding along the crease lines, a perforation pattern is locally applied at each crease. The perforation geometry used in this work is illustrated schematically in Fig.~\ref{fig:setup}A(i) and is defined by a set of geometric parameters, denoted by $C_1,\ldots,C_4$, which control the length, spacing, and width of the perforations. In all fabricated samples, these parameters are fixed and set to
$C_1 = C_2 = \SI{1}{mm}, \ 
C_3 = \SI{2}{mm}, \
C_4 = \SI{6}{mm}.$
These perforation parameters are used consistently for all tubular origami structures reported in this study.

\paragraph{Experimental characterization.}
We characterize the mechanical response of the fabricated tubular origami structures through a series of force--displacement experiments designed to probe their stiffness under different loading modes. All experiments are conducted using a tensile testing device equipped with force and moment sensors.

In Fig.~\ref{fig:setup}B(i), the tubular origami is mounted vertically between two clamps, with its axial direction aligned with the $z$ axis. In this setup, axial translational stiffness is measured by applying controlled displacements along the $z$ direction, and torsional stiffness about the $z$ axis is characterized by applying controlled rotation using the twist driver.

To characterize shear and bending responses, the experimental setup is modified as shown in Fig.~\ref{fig:setup}B(ii). We integrated a custom-designed rotating platform into the tensile testing machine, allowing the tubular origami to be positioned horizontally along its length. In this configuration, translational stiffness along the $x$ direction is measured by applying controlled displacements, corresponding to shear deformation of the tubular origami. In addition, rotational stiffness about the $x$ axis is measured by applying controlled rotations, probing the bending response of the structure. The rotating platform consists of two identical circular stages mounted at the ends of the origami structure. The left platform is connected to the movable crosshead of the tensile testing machine, while the right platform is connected to the force and moment sensors. By synchronously rotating both platforms to prescribed orientations about the $z$ axis and performing force--displacement measurements, the shear and bending stiffnesses can be measured along arbitrary directions within the $x$--$y$ plane.

Together, these experiments provide a comprehensive mechanical characterization of the critical stiffness responses of the fabricated origami structures.

\subsection{Numerical analysis}

We employ a general nonlinear formulation for the structural analysis of tubular origami based on a bar-and-hinge model \citep{liu2017nonlinear, filipov2017bar}. The formulation is displacement-based, with nodal displacements taken as the primary unknowns, from which rotations, internal forces, and stiffness are derived. Geometric nonlinearities are fully accounted for, enabling the analysis of the large deformations characteristic of origami structures. The bar-and-hinge system is assumed to be conservative, such that the total potential energy depends only on the current configuration and is independent of the deformation history. Equilibrium configurations are obtained by enforcing the principle of stationary potential energy, from which the nonlinear equilibrium equations and the corresponding tangent stiffness matrix are derived while consistently accounting for geometric nonlinearities. 

The strain energy of the structure is decomposed into two contributions: the energy stored in the bar elements and the energy stored in the rotational springs. Bar elements are placed along crease lines and across panels to capture axial and in-plane stiffness, respectively. Rotational hinges are assigned to bars along crease lines to model crease folding, and to bars spanning panels to represent panel bending behavior. In this model, the axial stiffness of bar elements is defined as $K_b=EA_e/L_e$, where $E = \SI{2}{GPa}$ is the Young's modulus of the paper, $L_e$ is the bar length, and $A_e$ is the effective bar area \cite{filipov2017bar}, representing the in-plane stretching resistance. The rotational stiffness of a crease hinge, $K_c = L_c k_c$, scales linearly with the crease length $L_c$, where $k_c = \SI{0.03}{N/rad}$ denotes the bending stiffness per unit length obtained from experimental tests on crease specimens (see Supplementary Material). For panel bending, the rotational hinges spanning the panels are assigned a stiffness $K_p = L_p k_p$, where $L_p$ is the length of the longest diagonal of the panel and $k_p$ denotes the panel bending stiffness per unit length. The value of $k_p$ is identified by fitting the numerical model to the two experimentally measured stiffnesses of Design I: the axial translational stiffness along the $z$-axis, $K_z = \SI{10}{N/m}$, and the torsional stiffness about the $z$-axis, $K_{rz} = \SI{11.5}{Nm}$. This yields $k_p = \SI{0.61}{N/rad}$, which is then kept fixed in all subsequent simulations throughout the paper.

We next compare the calibrated numerical model with experiments on the fabricated Design I. Fig. \ref{fig:exp_verify}A and B show the axial force--displacement response along the $z$-axis and the torsional response about the $z$-axis, respectively. These two responses are used to identify $k_p$. The remaining results are then used to assess the predictive capability of the model for stiffness responses not included in the fitting procedure.

Next we examine the in-plane translational stiffness of the tubular origami structure. In Fig.~\ref{fig:exp_verify}C(i), we show the directional translational stiffness evaluated over all orientations within the $x$--$y$ plane. 
The stiffness components $K_x$ and $K_y$ are computed at multiple prescribed axial displacements $D_z$, allowing to quantify how the in-plane constraint stiffness evolves as the structure deforms along its compliant $z$-direction.
The results indicate that the in-plane stiffness varies only moderately with axial deformation. The minimum stiffness occurs at a positive axial displacement of $D_z = \SI{5}{mm}$, corresponding to an extended configuration, where the combined in-plane stiffness reaches $K_{xy} = \SI{15.5}{N/m}$. Despite this relatively weak dependence on axial deformation, the stiffness exhibits pronounced directional anisotropy. The stiffness along the $x$-direction (approximately $\SI{25.1}{N/m}$) is significantly higher than that along the $y$-direction (approximately $\SI{15.9}{N/m}$). The directional dependence at zero axial displacement is shown explicitly in Fig.~\ref{fig:exp_verify}C(ii), where the magnitude of the in-plane translational stiffness is plotted as a function of the orientation angle. The numerical predictions are in close agreement with the experimental measurements.

We next analyze the rotational stiffness about axes lying in the $x$--$y$ plane. In Fig.~\ref{fig:exp_verify}D(i), we present the directional rotational stiffness components $K_{rx}$ and $K_{ry}$ evaluated at different axial displacements. In contrast to the translational response, the rotational stiffness is strongly influenced by axial deformation. The extended configuration (positive $D_z$) exhibits increased rotational stiffness, whereas compression along the negative $z$-direction reduces the stiffness. The minimum rotational stiffness is observed in the compressed configuration, reaching $K_{rxy} = \SI{0.11}{Nm}$. The directional dependence at zero axial displacement is shown in Fig.~\ref{fig:exp_verify}D(ii), where the rotational stiffness magnitude is plotted as a function of the axis orientation. Here, $K_{ry}$ (approximately $\SI{0.48}{Nm}$) exceeds $K_{rx}$ (approximately $\SI{0.17}{Nm}$) both from numerical predictions and the experimental results.

\subsection{Optimization} 
Having validated the nonlinear bar-and-hinge model against experiments (Design~I), we next use it as a predictive tool to search for tubular origami geometries that are compliant along the axial $z$-direction while remaining stiff in all other translational and rotational directions. For each generated topology (Fig.~\ref{fig:generation}), we compute stiffness quantities under large deformation by prescribing an axial displacement $D_z$ and evaluating the directional stiffness components at the corresponding deformed equilibrium configuration (Fig.~\ref{fig:exp_verify}). This procedure yields (i) the axial translational stiffness $K_z$ and the in-plane translational stiffness set $K_{xy}(\theta)$, and (ii) the torsional stiffness about the $z$-axis $K_{rz}$ and the torsional stiffness about axes lying in the $x$--$y$ plane $K_{rxy}(\theta)$, where $\theta$ denotes the in-plane orientation.

To quantify the desired anisotropic stiffness, high stiffness in all constrained directions relative to the compliant axial direction, we introduce two objective functions that are based on the weakest constrained direction. Specifically, we define the minimum in-plane translational stiffness as
\begin{equation}
K_{xy}^{\min}=\min_{\theta}\,K_{xy}(\theta),    
\end{equation}
and the minimum rotational stiffness among all non-compliant rotational modes as
\begin{equation}
K_{r}^{\min}=\min\!\left(\min_{\theta}\,K_{rxy}(\theta),\,K_{rz}\right).
\end{equation}

We then define two dimensionless objective functions,
\begin{equation}
\label{eq:objfun}
f_1 = \frac{K_z}{K_{xy}^{\min}}, 
\qquad
f_2 = \frac{K_{r_0}}{K_{r}^{\min}}.
\end{equation}
where $f_1$ quantifies translational anisotropy and $f_2$ quantifies rotational anisotropy. The constant $K_{r_0}= \SI{1}{Nm}$ is introduced as a reference rotational stiffness to nondimensionalize $f_2$ and to keep the two objective functions on comparable numerical scales.

Minimizing $f_1$ enforces a small ratio between the compliant axial stiffness and the weakest in-plane translational stiffness, thereby promoting structures that are compliant in $z$ while stiff in every in-plane direction. Likewise, minimizing $f_2$ promotes designs with large rotational stiffness in all constrained rotational directions by penalizing the minimum value among $\{K_{rxy}(\theta)\}$ and $K_{rz}$. In Fig.~\ref{fig:exp_verify}C(i) and Fig.~\ref{fig:exp_verify}D(i), the minima over $\theta$ correspond to the weakest in-plane translational and rotational directions (marked by red stars), which govern the two objectives.

Finally, we solve the resulting multi-objective design problem using the Non-dominated Sorting Genetic Algorithm~II (NSGA-II), which identifies a set of Pareto-optimal solutions balancing the two competing objectives. In each generation, candidate designs are (i) generated through the geometric parameterization (Fig.~\ref{fig:generation}), (ii) analyzed using the validated nonlinear bar-and-hinge model under the prescribed axial displacement $D_z$, (iii) assigned objective values $(f_1,f_2)$ according to \eqref{eq:objfun}, and (iv) evolved through selection, crossover, and mutation. The output of the optimization is a Pareto front of non-dominated tubular origami designs that achieve strong anisotropic stiffness properties.

\section{Results and Discussion}
\begin{figure*}[b!]
    \centering
    \includegraphics[width=\linewidth]{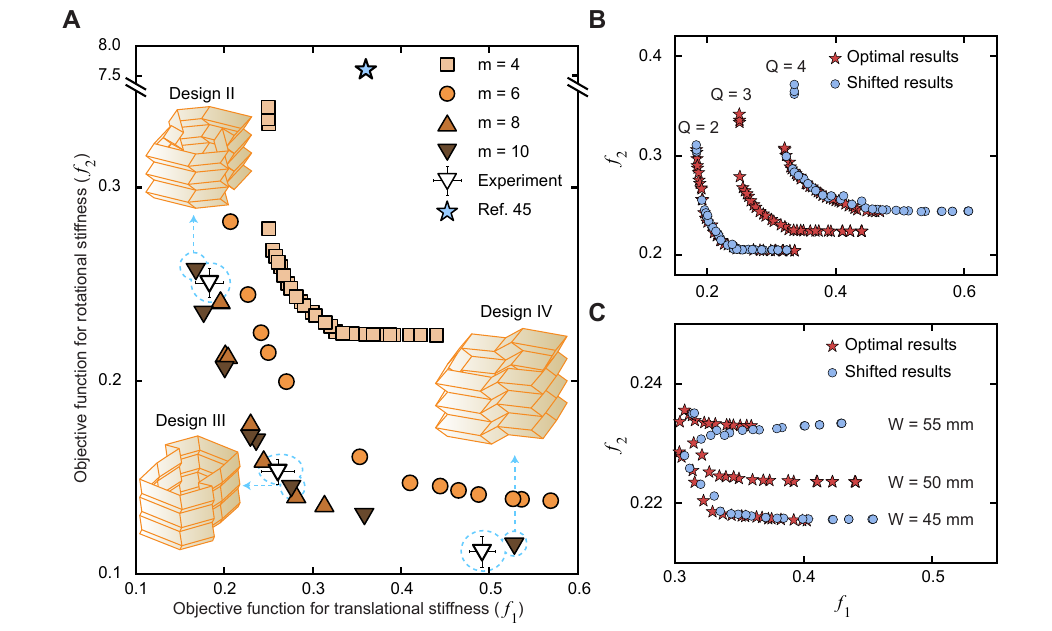}
    \caption{\textbf{Pareto fronts illustrating the influence of loop topology and geometric parameters on stiffness anisotropy.} (A) Pareto fronts for different numbers of centerline vertices $m$ (degree-4 vertex topology fixed). White markers indicate experimentally tested designs. (B) Pareto fronts for different numbers of stacked layers $Q$ (red: independently optimized; blue: designs optimized at $Q=3$ and re-evaluated for $Q=2$ and $Q=4$). (C) Pareto fronts for different layer heights $W$ (red: independently optimized; blue: designs optimized at $W=\SI{50}{mm}$ and re-evaluated at $W=\SI{45}{mm}$ and $W=\SI{55}{mm}$ without re-optimization).}
    \label{fig:PF}
\end{figure*}
\subsection{Influence of centerline loop topology}
We first examine the influence of loop topology, characterized by the number of vertices $m$ defining the closed polygonal centerline, on the achievable stiffness anisotropy. To isolate the effect of loop topology, the vertex topology is fixed to degree-4 vertices. Accordingly, each vertex is defined by a single projected crease vector $Z_{i,1}$ following the mirror-symmetric parameterization introduced earlier.

In Fig.~\ref{fig:PF}A, we show the Pareto fronts obtained for $m=4$, $6$, $8$, and $10$. Since both objective functions $f_1$ and $f_2$ are formulated such that smaller values correspond to stronger anisotropic stiffness performance, optimal designs lie toward the lower-left region of the plot. As $m$ increases, the attainable Pareto front progressively shifts toward lower objective values, indicating that increasing the number of polygonal vertices enriches the admissible geometric space and enables improved stiffness trade-offs. Notably, the Pareto fronts corresponding to $m=8$ and $m=10$ nearly overlap in the optimal region, suggesting convergence with respect to loop topology refinement. Increasing $m$ beyond 8 therefore provides marginal performance improvement. To further analyze the mechanical implications of loop topology, three representative optimized designs with $m=10$ were selected from distinct regions of the Pareto front, referred to as Design II, Design III, and Design IV.
Design~II is defined by polygon vertices 
$P_i = (x_i,y_i)$, $i=1,\ldots,10$, with 
$P_1=(15.0,17.3)$, $P_2=(37.2,15.0)$, $P_3=(50.8,23.2)$, $P_4=(40.6,42.6)$, $P_5=(48.3,38.2)$, $P_6=(50.0,57.7)$, $P_7=(27.8,60.0)$, $P_8=(14.2,51.8)$, $P_9=(24.4,32.4)$, $P_{10}=(16.7,36.8)$,
and projected crease vectors 
$Z_{i,1}=(2.7,4.2)$ for $i=1,\ldots,10$. 
This design achieves the lowest translational objective value among the three, indicating strong in-plane translational stiffness relative to axial compliance, while maintaining moderate rotational performance.
Design~III is defined by $P_1=(15.0,17.3)$, $P_2=(38.1,9.5)$, $P_3=(49.7,22.9)$, $P_4=(42.3,23.7)$, $P_5=(47.5,40.3)$, $P_6=(50.0,57.7)$, $P_7=(26.9,65.5)$, $P_8=(15.3,52.1)$, $P_9=(22.7,51.3)$, $P_{10}=(17.5,34.7)$ and $Z_{i,1}=(-2.6,4.2)$ for $i=1,\ldots,10$. It lies near the center of the Pareto front and represents a balanced trade-off between translational and rotational anisotropy.
Design~IV is defined by 
$P_1=(15.0,17.3)$, $P_2=(33.3,17.5)$, $P_3=(37.0,26.6)$, $P_4=(50.6,28.3)$, $P_5=(50.7,43.5)$, $P_6=(50.0,57.7)$, $P_7=(31.7,57.5)$, $P_8=(28.0,48.4)$, $P_9=(14.4,46.7)$, $P_{10}=(14.3,31.5)$ and $Z_{i,1}=(3.5,3.5)$ for $i=1,\ldots,10$. 
This configuration achieves the lowest rotational objective value $f_2$, corresponding to enhanced rotational stiffness in constrained directions, while preserving acceptable translational anisotropy.

All three geometries were fabricated and experimentally characterized. The experimentally determined objective-function values (white markers in Fig.~\ref{fig:PF}A) agree with the numerical predictions to within $8\%$ for all three designs.

For comparison, we evaluated the origami design reported in Ref.~\citep{filipov2015origami-pnas-tube} using the same objective functions and under equivalent conditions. The reference design gives $f_1 = 0.36$ and $f_2 = 7.60$, whereas the optimized topology obtained here (Design~III) gives $f_1 = 0.28$ and $f_2 = 0.15$. Relative to the reference design, Design~III reduces $f_1$ by a factor of $1.3$ and $f_2$ by a factor of $50.7$. Since $f_2$ is inversely proportional to the minimum constrained rotational stiffness, this corresponds to a minimum constrained rotational stiffness that is $50.7$ times higher than that of the reference design.

We next investigate the influence of other geometric parameters, namely the number of stacked layers $Q$ and the layer height $W$, while keeping the loop topology fixed. Fig~\ref{fig:PF}B shows the Pareto fronts obtained by independently optimizing the structures for $Q=2$, $3$, and $4$ (red markers). To assess the sensitivity of the optimized topologies to the number of layers, we take the optimal designs obtained for $Q=3$ and re-evaluate them after changing only the layer number to $Q=2$ and $Q=4$, without performing a new optimization. The resulting objective values are shown as blue markers. The close overlap between the independently optimized fronts (red markers) and the re-evaluated designs (blue markers) indicates that the optimized topologies are largely insensitive to the number of layers. Changing $Q$ primarily induces a uniform shift in the objective values, while preserving the relative position and curvature of the Pareto front. This demonstrates that the translational–rotational stiffness trade-off is governed predominantly by loop topology, and is only weakly affected by the stacking number.

In Fig.~\ref{fig:PF}C, we show a similar robustness analysis for variations in the layer height $W$. Red markers correspond to independently optimized Pareto fronts for $W=45$, $50$, and $\SI{55}{mm}$. To assess sensitivity to geometric scaling, the optimized designs obtained for $W=\SI{50}{mm}$ are re-evaluated after modifying only the layer height to $W=\SI{45}{mm}$ and $W=\SI{55}{mm}$, without re-optimization (blue markers). The close alignment between the independently optimized and re-evaluated results indicates that varying $W$ primarily induces a uniform shift in the objective values while preserving the shape and relative ordering of the Pareto front.

Taken together, these results indicate that loop topology is the dominant factor controlling stiffness anisotropy, while geometric parameters such as $Q$ and $W$ mainly modulate the stiffness magnitude without substantially altering the trade-off structure.

\begin{figure*}[t!]
    \centering
    \includegraphics[width=\linewidth]{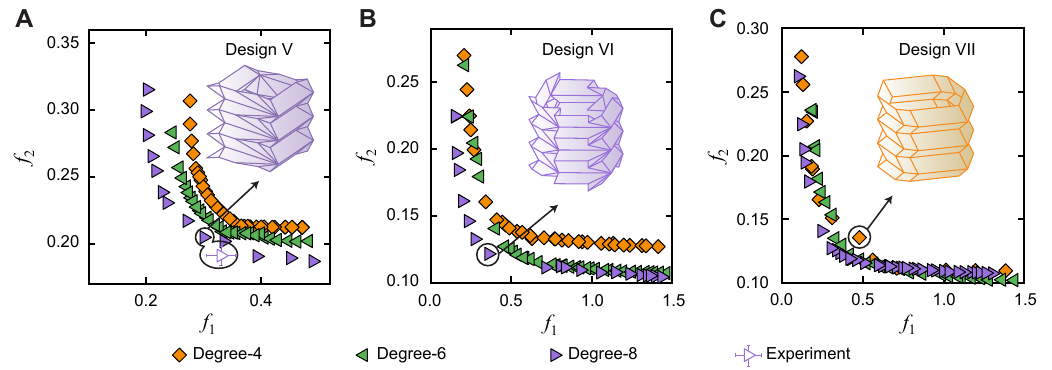}
    \caption{\textbf{Pareto fronts showing the influence of vertex degree on stiffness anisotropy.} Pareto fronts for degree-4, degree-6, and degree-8 vertices evaluated at fixed centerline vertex numbers (A) $m=4$, (B) $m=6$, and (C) $m=8$.}
    \label{fig:vertex}
\end{figure*}

\subsection{Influence of vertex topology}
We next investigate the influence of vertex topology by varying the vertex degree ($n=4, 6, 8$) while fixing the number of centerline vertices $m$. In Fig.~\ref{fig:vertex}A–C, the Pareto fronts for each vertex degree are evaluated separately for $m=4$, $m=6$, and $m=8$, respectively.

For $m=4$ (Fig.~\ref{fig:vertex}A), increasing the vertex degree systematically improves stiffness anisotropy. The Pareto front corresponding to degree-8 vertices extends toward lower values of both objective functions compared to degree-4 vertices, indicating enhanced translational and rotational stiffness performance. Degree-6 vertices provide intermediate performance, particularly improving torsional stiffness relative to degree-4 configurations. This demonstrates that when the centerline loop topology is limited ($m=4$), increasing vertex degrees of freedom significantly enlarges the attainable design space. A representative optimized configuration for this case is shown in Fig.~\ref{fig:vertex}A (Design~V), which employs degree-8 vertices with $m=4$. The geometric parameters are defined by $P_1=(15.0,17.3)$, $P_2=(10.7,54.9)$, $P_3=(50.0,57.7)$, $P_4=(54.3,20.1)$ and projected crease vectors $Z_{1,1}=(-3.9,-0.9)$, $Z_{1,2}=(-8.9,-6.0)$, $Z_{1,3}=(-6.2,1.8)$, $Z_{2,1}=(-5.0,-8.4)$, $Z_{2,2}=(5.1,-8.7)$, $Z_{2,3}=(18.7,-4.0)$, $Z_{3,1}=(-8.9,-6.0)$, $Z_{3,2}=(-3.9,-0.9)$, $Z_{3,3}=(-6.6,-8.7)$, $Z_{4,1}=(-7.7,1.5)$, $Z_{4,2}=(-17.8,1.7)$, $Z_{4,3}=(-31.4,-2.9)$. We fabricated and experimentally characterized Design V, with the measured objective values deviating by less than $7\%$ from the numerical predictions.

For $m=6$ (Fig.~\ref{fig:vertex}B), a similar trend is observed: higher vertex degree leads to improved Pareto performance. Degree-8 vertices again achieve the lowest attainable objective values, although the relative improvement compared to lower-degree vertices becomes less pronounced than in the $m=4$ case. A representative optimized configuration (Design~VI) is shown in Fig.~\ref{fig:vertex}B, defined by $P_1=(15.0,17.3)$, $P_2=(3.6,42.4)$, $P_3=(29.1,63.1)$, $P_4=(50,57.7)$, $P_5=(61.4,32.6)$, $P_6=(35.9,11.9)$ and corresponding vectors $Z_{1,1}=(-5.3,-0.4)$, $Z_{1,2}=(-10.3,2.9)$, $Z_{1,3}=(-6.5,-4.7)$, $Z_{2,1}=(-5.4,2.2)$, $Z_{2,2}=(-5.2,15.2)$, $Z_{2,3}=(0.8,8.0)$, $Z_{3,1}=(-8.1,-0.7)$, $Z_{3,2}=(-4.7,-9.6)$, $Z_{3,3}=(5.1,-3.1)$, $Z_{4,1}=(-5.3,-0.4)$, $Z_{4,2}=(-0.4,-3.7)$, $Z_{4,3}=(-4.2,-5.5)$, $Z_{5,1}=(-5.3,-3.0)$, $Z_{5,2}=(-5.4,-15.9)$, $Z_{5,3}=(-11.5,-8.7)$, $Z_{6,1}=(-2.6,-1.5)$, $Z_{6,2}=(-6.0,8.9)$, $Z_{6,3}=(-15.8,2.3)$. Design~VI achieves $f_1=0.35$ and $f_2=0.12$, which implies a $63.3$-times higher minimum constrained rotational stiffness than the benchmark design reported in Ref.~\cite{filipov2015origami-pnas-tube}.

For $m=8$ (Fig.~\ref{fig:vertex}C), the Pareto fronts corresponding to degree-4, degree-6, and degree-8 vertices nearly coincide in the optimal region. Increasing the vertex degree in this case provides only marginal improvement. A representative configuration (Design~VII) is shown in Fig.~\ref{fig:vertex}C with geometry defined by $P_1=(15.0,17.3)$, $P_2=(4.2,35.5)$, $P_3=(22.7,60.0)$, $P_4=(30.6,64.2)$, $P_5=(50.0,57.7)$, $P_6=(60.8,39.5)$, $P_7=(42.3,15.0)$, $P_8=(34.4,10.8)$ and projected vectors $Z_{i,1}=(-8.2,1.6)$ for $i=1,\ldots,8$. These results indicate that vertex topology plays a significant role when the loop topology is limited (small $m$), but its influence diminishes as the number of centerline vertices increases. For sufficiently large $m$, loop topology dominates the attainable stiffness anisotropy, and increasing vertex degree yields diminishing returns.

\subsection{Uncertainty analysis}
To evaluate the sensitivity of the optimized designs to geometric imperfections, we perform a Monte Carlo analysis on the $m=4$ designs located on the Pareto front in Fig.~\ref{fig:PF}A. Only uncertainties associated with the polygon vertices $P_i$, which define the centerline loop, are considered.

\begin{figure}[t!]
    \centering
    \includegraphics[width=\columnwidth]{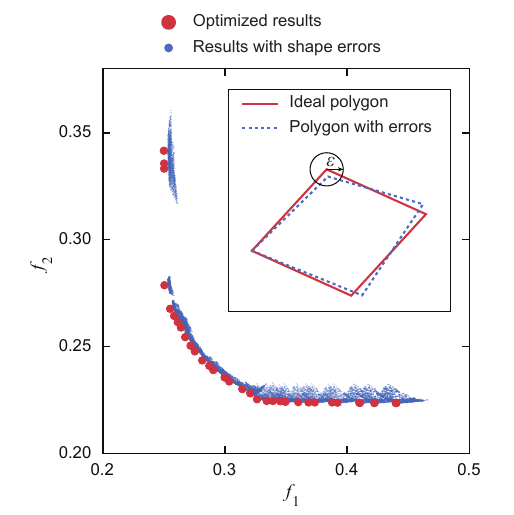}
    \caption{\textbf{Sensitivity of optimized designs to geometric imperfections.} Red markers denote optimized designs on the Pareto front of the degree-four vertex with $m=4$, while blue markers show results from Monte-Carlo simulations with random perturbations applied to the polygon vertices. Inset: schematic of the ideal polygon (solid) and a perturbed polygon (dashed), where each vertex is randomly displaced within a radius $\varepsilon = 1$ mm.}
    \label{fig:uncert}
\end{figure}

The polygon defined by the optimized geometries is referred to as the \textit{ideal polygon} (red lines in Fig.~\ref{fig:uncert}). Geometric imperfections are introduced by applying small random perturbations to each vertex, producing perturbed vertices $P_i'$. The resulting geometry is referred to as the \textit{polygon with errors} (blue dashed lines in Fig.~\ref{fig:uncert}). The perturbations are constrained by
\begin{equation}
\sqrt{(\Delta x)^2 + (\Delta y)^2} < \varepsilon,
\end{equation}
where $\Delta x$ and $\Delta y$ are random in-plane displacements and $\varepsilon = 1~\mathrm{mm}$ defines the maximum allowable deviation. 
%Consequently, each perturbed vertex lies within a disk of radius $\varepsilon$ centered at the original position.
For each optimized design, 500 random realizations are generated and evaluated numerically. The performance of the ideal designs is shown by red markers in Fig.~\ref{fig:uncert}, while the perturbed realizations are shown in blue.

The results indicate that geometric imperfections produce only minor deviations from the optimal performance, with slightly increased objective values compared to the ideal geometries. This demonstrates that the optimized designs lie near the true Pareto boundary and retain robustness against small geometric errors.

% The original one:
% To examine the deterioration of performance resulting from small deviations in the design, Monte-Carlo simulations were run. Small errors were introduced in each vertex of the centerline loop ($P_{i}$) of the best design (Fig. \ref{fig:PF}(a)) \DF{Which design are you referring to? Give the name here as we introduced earlier.} For each of these vertices, a random error in the $x$ and $y$-direction is added of maximum size $\sqrt {\Delta {x^2} + \Delta {y^2}}  < \varepsilon $ \DF{I do not understand this you need to explain it better. what was the maximum and minimum value of the error you allow? what is the $\varepsilon$ and how big is it?}. The size of shape errors for each vertex is the same, and 1000 random variations are analyzed. This resulted in the data shown in Fig. \ref{fig:uncert}. The Pareto front consists of red dots representing optimal results \DF{which results are these? a new pareto front? or these are the same as one of the earlier pareto front? You need to better explain this!}. Results with errors are indicated by blue dots. It can be seen from the figure that after the error is introduced, the result will have a small deviation in the direction of stiffness weakening \DF{cant follow this explanation, what does "....in the direction of stiffness weakening..." means?}, which can explain: 1. Our optimization algorithm has found the optimal result; 2. The optimized structure has a certain robustness.

\section{Conclusions}
We introduced an automated design framework for tubular origami that explores both local vertex topology and global tube topology through the polygonal cross-section. Using a nonlinear bar-and-hinge model, objective functions were formulated to evaluate translational and rotational anisotropic stiffness under large deformation, and Pareto fronts were used to identify optimal designs. The results show that increasing the number of vertices in the polygonal cross-section improves the optimization results, although the improvement becomes smaller as the number of cross-sectional vertices increases. Increasing the vertex degree further improves performance, particularly for tubes with a small number of vertices in the polygonal cross-section, showing that higher local kinematic freedom does not necessarily compromise stiffness at the structural scale. Compared with the state-of-the-art design, the optimized structures achieve more than $50$ times higher constrained rotational stiffness, which was supported by experiment. These results identify polygonal cross-sectional topology and higher-degree vertices as effective design variables for programming anisotropic stiffness in tubular origami. Future work can build on this framework by, for example, solving the inverse problem of finding tube geometries that realize prescribed stiffness characteristics for desired deformation modes, or by exploiting the additional local degrees of freedom in higher-degree vertices for reconfiguring and modulating stiffness properties post fabrication.

%In this work, a design method for origami tubes was proposed. With the help of middle line trajectory and geometry parameters, various origami tubes were generated. Objective functions for translational stiffness and torsional stiffness in all constrained translational and rotational directions were formulated to evaluate the mechanical properties of origami tubes when considering large deformations. The Pareto front was obtained to evaluate the performance of the structure in both translation and torsion. By increasing the number of middle-line trajectory vertices, the optimization results can be significantly improved at the beginning. But when $m>10$, the gain of increasing $m$ will become smaller. On the other hand, the degree of freedom of vertices has a positive effect on obtaining better structures when $m$ is small ($m<6$). The two geometric parameters ($Q$ and $W$) that determine the extension in the out-of-plane direction will not affect the tendency of the structure to translation and torsion. The results were verified experimentally. The proposed design framework for origami tubes can be potentially useful in various engineering applications.

\acknow{}
\showacknow{} % Display the acknowledgments section
The authors acknowledge financial support from ASML and TKI HTSM. The authors also thank Frido van der Blij, Luis Garcia, and Jelle Rommers from ASML for valuable technical discussions.% Bibliography
% \nocite{*} % show all references in bib file
% \bibliography{bib_file}

\section*{REFERENCES}
\bibliography{MyRef}

\end{document}

%% Comment out or remove this line before generating final copy for submission; this will also remove the warning re: "Consecutive odd pages found".
% \instructionspage  

\maketitle

%% Adds the main heading for the SI text. Comment out this line if you do not have any supporting information text.
\SItext

\section{Measurement of rotational stiffness of creases}
\label{sec:stiffness of crease}
\begin{figure*}[htb!]
    \centering
    \includegraphics[width=\linewidth]{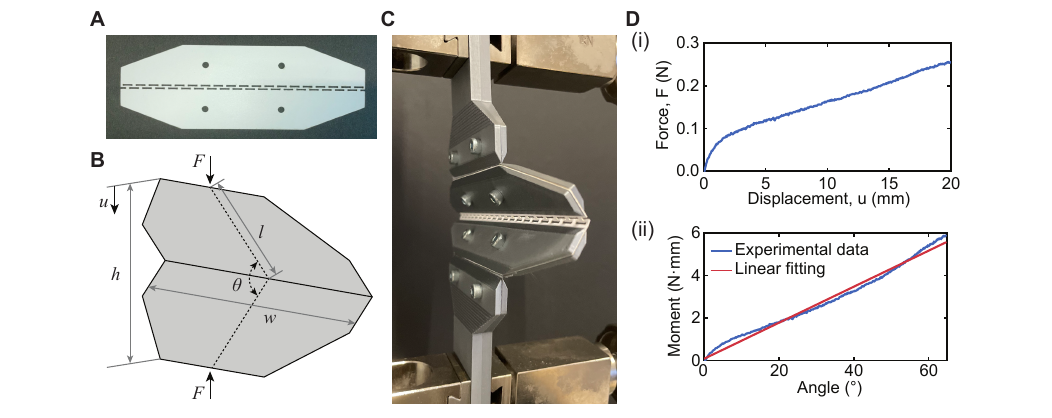}
    \caption{\textbf{Rotational stiffness testing for crease hinges.} (A) A crease hinge sample manufactured with laser cutting. (B) A schematic diagram and parameter definitions for a crease hinge. (C) Experimental setup used for rotational stiffness. (D) Force-displacement relationship of crease hinges under uni-axial compression test.}
    \label{fig:crease}
\end{figure*}

In order to determine the rotational stiffness of creases, a uniaxial compression test is performed on an origami single-crease hinge. The crease hinge sample, as shown in Fig.~\ref{fig:crease}A, is fabricated using laser cutting. The crease line is located at the center of the specimen. The upper and lower panels, separated by the crease line, are clamped by rigid plates using screws. Therefore, it is assumed that the two panels remain flat and rigid, and deformation occurs only along the crease line.

The deformed configuration of the crease hinge is illustrated schematically in Fig.~\ref{fig:crease}B. The width of the crease hinge is $w = \SI{160}{mm}$, and the length of each panel is $l = \SI{27.5}{mm}$. Under a compressive force $F$, the hinge undergoes a displacement $u$. The moment applied to the crease hinge can be expressed as $M = F l \cos(\theta/2)$. The dihedral angle is given by $\theta = 2 \arcsin\left(\frac{h}{2l}\right)$.

The force–displacement relationship is obtained from the experimental setup shown in Fig.~\ref{fig:crease}C.
The rotational stiffness is extracted from the experimentally measured force–displacement data (Fig.~\ref{fig:crease}D(i)). The displacement is first converted to the dihedral angle based on the geometric relation, and the applied force is transformed into the bending moment. This yields the moment–rotation relationship, as shown by the blue line in Fig.~\ref{fig:crease}D(ii). The data is then fitted using a linear function $M = k\theta$ as shown by the red line in Fig.~\ref{fig:crease}D(ii). The rotational stiffness per unit length is subsequently obtained from the slope of the fitted curve as $k_c = k/w$.

%%% Add this line AFTER all your figures and tables
\FloatBarrier

% Bibliography
% \nocite{*} % show all references in bib file
% \bibliography{bib_file}